\documentclass[aps,prb,twocolumn,showpacs,floatfix]{revtex4}
\usepackage{graphicx}

\begin{document}

\newcommand{\mn}{MnV$_2$O$_4\,\,$}
\newcommand{\lc}{\lowercase}

\title{Proposed Orbital Ordering in MnV$_2$O$_4$ from First-principles Calculations}

\author{S. Sarkar$^{1}$, T. Maitra$^{2}$, Roser Valent{\'\i}$^{3}$, T. Saha-Dasgupta$^{1}$}

\affiliation{ $^{1}$ S. N. Bose National Centre for Basic Sciences, Kolkata, India\\
$^{2}$ Department of Physics, Indian Institute of Technology, Roorkee, India\\
$^{3}$Institut f{\"u}r Theoretische Physik, J. W. Goethe Universit{\"a}t, Frankfurt, Germany}

\date{\today}

\begin{abstract}
Based on density functional calculations, we propose a possible orbital ordering in MnV$_2$O$_4$
which consists of orbital chains running along crystallographic $a$ and $b$ directions with orbitals rotated alternatively
by about 45$^{\circ}$ within each chain.
 We show that the consideration of correlation
effects as implemented in the local spin density
approximation (LSDA)+U approach is crucial for a correct
description of the space group symmetry.  This implies that
 the correlation-driven orbital ordering has a strong influence
on the structural transitions
in this system. Inclusion
of spin-orbit effects does not seem to influence the orbital ordering pattern. We further
find that the proposed orbital arrangement favours a noncollinear magnetic ordering of V spins,
as observed  experimentally. Exchange  couplings among
 V spins are also calculated and discussed.

\footnotesize{}

\end{abstract}
\maketitle

The spinel compounds with a chemical formula of AB$_2$X$_4$ where B sites are usually transition metal ions,
form a frustrated pyrochlore lattice with corner-sharing tetrahedra. These compounds show
a complex behavior including   structural transitions
from cubic to tetragonal symmetries which are
often accompanied by an orbital order-disorder transition as well as complicated magnetic orderings
at low temperatures\cite{radaelli}.

The spinel \mn	has experienced a recent surge in activities due
to new	experimental observations in single crystals\cite{garlea} revealing a lower symmetry structure than previously
suggested\cite{adachi}. This has important implications for the related orbital order at
low temperatures which is still unclear. The presence of two magnetic ions in \mn
(Mn with spin 5/2 and V with spin 1) translates into
 more complex magnetic phase transitions
in this system than in other vanadium spinel oxides such as ZnV$_2$O$_4$, MgV$_2$O$_4$ or CdV$_2$O$_4$ with
nonmagnetic A-site ions. Recent experimental findings\cite{garlea,suzuki} indicated that \mn
undergoes a phase transition from paramagnetic to a collinear ferrimagnetic
phase at 56K
where the Mn and V spin moments point in opposite directions. At T = 53K
a second magnetic phase transition to noncollinear ferrimagnetism
follows	 accompanied by a
structural transition from cubic to tetragonal phase.

The  cubic to tetragonal structural transition in \mn is, similar to other vanadium spinels,
associated with a compression of the VO$_6$ octahedron ($c_T/a_T=0.98$).
The octahedral environment  of V  (VO$_6$)  splits the $d$ states into lower t$_{2g}$
and higher e$_g$.  Since  V$^{+3}$ is in a 3$d^2$ configuration, the t$_{2g}$ orbitals are partially filled and possible orbital orderings may occur.
Earlier experimental observations \cite{adachi} indicated the tetragonal space group to be
$I4_1/amd$. However,
recent precise measurements on a single crystal\cite{garlea, suzuki}
showed that  the tetragonal space group is  $I4_1/a$.
Since  the orbital order and, accordingly,  the magnetic order are closely related to the
 underlying space group symmetry, it is very important	to
establish the space group symmetry unambiguously.

The $I4_1/a$ space group
breaks the mirror and glide symmetries present in the $I4_1/amd$ space group,
which implies that two of the four V-O bonds
in the $ab$ plane are shorter  whereas in $I4_1/amd$ symmetry
all four V-O bond lengths are  the same.
 Garlea {\it et al.}\cite{garlea}  proposed a staggered
A-type orbital ordering for this system
based on their	observations
of the structural and magnetic phases at low temperature. A similar ordering was
also  proposed by Suzuki {\it et al.}\cite{suzuki}. Though the magnetic structure at low temperatures
has been unambiguously established by the above mentioned experiments, there has not yet been
any experiment such as X-ray resonant spectroscopy to directly probe the orbital order.
Determination of exchange couplings using neutron scattering techniques by Chung
{\it et al.}\cite{chung}  is in apparent contradiction with the proposed staggered A-type orbital ordering.
As pointed out by these authors,  the proposed orbital order
in Refs.~\onlinecite{garlea,suzuki}
lacks the consideration of trigonal distortion, which is found to be  strongest in \mn among  all the vanadium spinels.
The trigonal distortion has often shown to have significant effects on the orbital order \cite{maitra, anisimov}.

In this Letter we show, based on density functional theory (DFT) calculations,
that the ground state tetragonal space group symmetry at low temperatures is   $I4_1/a$ and strongly
driven by correlation effects.	We propose
an orbital ordering consisting of orbital chains  running along the axes $a$ and $b$ with
orbitals rotated by about 45$^{\circ}$ within each chain.  This ordering favors a noncollinear
arrangement of spins, as observed experimentally, which is a convincing indication of its existence.

 For our DFT calculations we  considered a combination of three
different methods, namely: (a) plane wave-based method (b) linear augmented plane wave (LAPW) method
and (c)	 muffin-tin orbital (MTO) based N-th order MTO (NMTO) method. Results
were cross-checked among the three schemes in terms of
 total energy differences, density of states and band structures.
Since first principles calculations take into account all
structural and chemical aspects appropriately, we expect to  gain a better
understanding of the nature of the structural phase transition and possible
orbital ordering.

We first performed a structural optimization using the plane wave method
as implemented in the Vienna Ab-initio Simulation Package (VASP)\cite{vasp} and
 considered exchange-correlation functionals within LSDA, generalized gradient
approximation (GGA) and LSDA+U\cite{Anisimov_93} in
order to investigate the  relative stability between $I4_1/amd$ and $I4_1/a$
symmetries in \mn .
We  used projector augmented wave (PAW) potentials\cite{paw} and the wavefunctions were
 expanded in the plane wave basis with a kinetic energy cut off of 450 eV. Reciprocal
space integration was carried out with a k-mesh of 6$\times$6$\times$6.

\begin{table}
\caption{Energy-minimized structural parameters for \mn. Lattice constants were
 kept  at the experimental value\cite{adachi}. The LSDA+U optimized
structural parameters  show the	 O x-coordinate	 to be
 non-zero, signaling the change of space group symmetry to $I4_1/a$.}
\begin{tabular}{|cc|c|c|}
\hline
& LSDA & GGA & LSDA+U  \\
&     &	    & (U=4.5 eV) \\
\hline
Mn & 0.0 0.25 0.125  & 0.0 0.25 0.125 &	 0.0 0.25 0.125 \\
V  & 0.0 0.0 0.5 & 0.0 0.0 0.5	& 0.0 0.0 0.5\\
O  & 0.0 0.0243 0.7392 & 0.0 0.0236 0.7394 & 0.0059 0.0244 0.7383 \\ \hline
\end{tabular}
\end{table}

Optimization of the atomic positions\cite{note} within LSDA  as well as GGA assuming
ferrimagnetic spin arrangements	 between Mn
and V atoms
gave us a ground state structure of $I4_1/amd$ symmetry where the tetragonal distortion
is found to be substantially reduced compared to the experimental estimate\cite{adachi}. In order to
check the influence of electron-electron correlation on the structural optimization, which has been
found to be important in previous reports\cite{u-effect}, we have further optimized the
atomic positions within the LSDA+U approach with different choices of U values\cite{uvalue} (U=0.5, 1, 2, 3,
 4.5
and 6 eV) for both   Mn and V. $J$ was
chosen
 to be 1 eV for all
calculations. Remarkably, we observe that with the consideration of U beyond 2eV, the $I4_1/a$ symmetry becomes the
ground state structure (see Table I). This optimized structure shows a tetragonal distortion close to the experimentally reported
one\cite{adachi}. These results	 indicate the importance of
correlation effects for the description of the correct
orbital ordering and the low temperature structure.

\begin{figure}
\includegraphics[width=8cm]{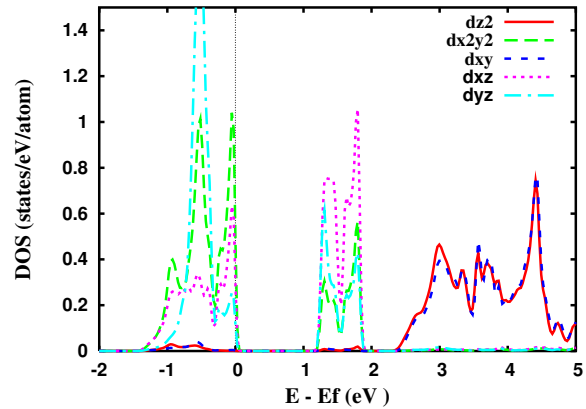}
\caption{(Color online) LSDA+U V-$d$ partial DOS
 for U=4.5 eV in the APW+lo basis. Only the DOS for the majority spin channel is shown
 (the minority spin channel is unoccupied).}
\label{dos}
\end{figure}


We analyzed the resulting orbital order with
the  full potential LAPW method as implemented in the Wien2k code
\cite{wien2k}.
The atomic sphere radii were chosen to be 2.01, 1.98 and 1.77 a.u. for
Mn, V and O respectively.   We chose the APW+lo as the basis set and the expansion in spherical
harmonics for the radial wave functions was taken up to $l=10$.
 The charge densities and
potentials were represented by spherical harmonics up to $l=6$. For Brillouin-
zone (BZ) integrations we considered a 52 $ k$ points mesh in the
irreducible wedge and the modified tetrahedron method was applied\cite{tetra}. The
collinear ferrimagnetic spin arrangements between Mn and V  was taken
the same as for
the structural optimization calculations.
In all further calculations we considered the LSDA+U approximation\cite{double} and fixed the value of
U at 4.5 eV which reproduces the experimentally observed orbital
moment in vanadium,
 as will be discussed later.

\begin{figure}
\includegraphics[width=9cm]{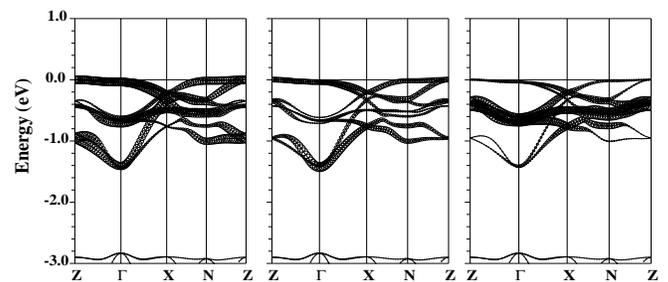}
\caption{LSDA+U bandstructure of \mn (APW+lo basis)
 projected onto V- x$^{2}$-y$^{2}$, xz and yz
character (from left to right) in the energy range [-3 eV, 1 eV].
The high symmetry path of the tetragonal Brillouin zone was considered.}
\label{bands}
\end{figure}

In Fig.~\ref{dos} we show the electronic density of states (DOS) calculated
within the LSDA+U approximation.
 In the partial DOS one observes the usual t$_{2g}$ (consisting of x$^{2}$-y$^{2}$,
xz and yz orbitals defined in the crystallographic co-ordinate system)\cite{note1} and e$_g$ (consisting of xy, 3z$^{2}$)
splitting of V $d$-orbitals due to the O octahedral crystal field.
Inclusion of correlation effects in the V $d$-orbitals through
the LSDA+U approach, splits the t$_{2g}$ states further and opens a gap of 1.1 eV. The degeneracy between all the three t$_{2g}$ orbitals is lifted
in the low symmetry $I4_1/a$ group\cite{comment}. All  t$_{2g}$ orbitals
 are  partially occupied
with higher x$^{2}$-y$^{2}$ and yz occupancy  compared to xz.
 This becomes more evident in
the band structure results.   Fig.~\ref{bands} shows the t$_{2g}$
bandstructure in the majority spin channel, which is separated from occupied O-p
dominated bands by a gap of  1.5 eV  and from unoccupied e$_{g}$-like bands
by a gap of  0.2 eV. The fatness of the bands indicate
the projected band characters
of x$^{2}$-y$^{2}$, xz and yz orbitals.

\begin{figure}
\includegraphics[width=6.5cm]{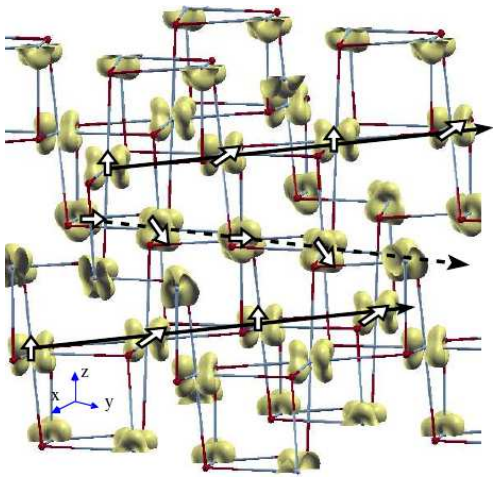}
\caption{{ (Color online) Three dimensional electron density plot showing the orbital ordering. The black
solid and dashed lines designate the orbital chains.  The arrows
superimposed on the electron density at each V site, mark the
rotation sense of the orbitals
as one moves to neighboring V sites within a given chain. The atoms at the alternate corners of the distorted cubes
are occupied by V and O respectively.
The isovalue was chosen as 0.1 e$^{-}$/$(\AA^{3})$.}}
\label{eldens}
\end{figure}

 Significant mixing of	orbitals
happens due to	the  low symmetry of the $I4_1/a$ space group.
In Fig.~\ref{eldens}  we show the three-dimensional electron density of
occupied V t$_{2g}$ orbitals on a real space grid.
  We identify a long range order pattern for the orbital distribution.
Contrary to the proposed  staggered A-type
order\cite{garlea,suzuki}, we observe orbital chains along $a$ and $b$ directions
(indicated by solid and dashed lines) with the orbitals within each chain rotated
alternatively by about 45$^{\circ}$  (shown by the arrows).

In order to assign the precise V orbital compositions
 we have performed NMTO-downfolding\cite{nmto} calculations
to construct a V-t$_{2g}$-e$_g$ only low-energy Hamiltonian by integrating out
 degrees of freedom other than V-t$_{2g}$-e$_g$, starting with a full LSDA+U Hamiltonian. Diagonalization of
the on-site energy block of this 5 x 5 Hamiltonian gives rise to eigenstates given by:

\begin{eqnarray*}
\vert 1 \rangle =\phantom{-}.78 \vert x^2-y^2 \rangle - .59 \vert xz \rangle - .21\vert yz \rangle
+ .07 \vert xy \rangle + .02   \vert z^2 \rangle\\
\vert 2 \rangle	 =-.35 \vert x^{2}-y^{2} \rangle
 - .15 \vert xz \rangle - .92 \vert yz
 \rangle - .09	\vert xy \rangle - .07	 \vert z^2 \rangle\\
\vert 3 \rangle	 =\phantom{-}.52 \vert x^{2}-y^{2} \rangle +
 .79 \vert xz \rangle - .31 \vert yz \rangle
- .13  \vert xy \rangle + .02	\vert z^2 \rangle\\
\vert 4 \rangle	 =\phantom{-}.05 \vert x^{2}-y^{2} \rangle - .08 \vert xz \rangle +
.11 \vert yz \rangle
- .66 \vert xy \rangle - .74   \vert z^2 \rangle\\
\vert 5 \rangle =-.02 \vert x^{2}-y^{2} \rangle - .11 \vert xz \rangle +
.04 \vert yz \rangle - .73 \vert xy \rangle + .67  \vert z^2 \rangle
\end{eqnarray*}
with energies 0.81, 1.19, 1.47, 2.05, 2.28 eV respectively.

We observe that the lowest energy state has predominant x$^{2}$-y$^{2}$ character -which is
expected due to the tetragonal distortion with
the compression of VO$_6$ octahedron along the {\it c}-direction-  with a
significant mixing of xz character.  The next higher energy
state is dominated by yz character. Therefore, the second electron of
$V^{3+}$  always occupies the orbital with predominant yz character in all V sites.
 The rotation of orbitals with respect
to each other within the chain and between the chains (see Fig.~\ref{eldens}),
 can therefore be explained due to the staggered trigonal distortion
that is present both within the $ab$-plane and along the {\it c}-direction.
Despite an apparent {\it antiferro-orbital} ordering, we call the
     ordering {\it ferro-orbital} since it is in all sites the
     same orbital that is occupied by the second electron, and
    not an alternating occupation of xz and yz.

The spin-orbit effect has been observed to play a significant role in dictating the nature of orbital
order\cite{tchernyshyov,maitra} in ZnV$_2$O$_4$ and was proposed to be
important
for   the magnetic and orbital physics of \mn\cite{plumier}.
We performed LSDA+U+SO calculations with the same U values as mentioned above,
where the spin-orbit effects have been introduced as a second variation using the
scalar relativistic approximation. Contrary
to the case of	ZnV$_2$O$_4$\cite{maitra}, we do not
observe any significant difference in charge density, from that of LSDA+U.
The value of the orbital moment depends sensitively on U. The experimental V moment
is best described for U=4.5 eV. At this U value we obtain an
orbital moment of about 0.34 $\mu_B$ at V site which is antiparallel to the spin-moment (1.65$\mu_B$). The total magnetic moment of 1.31 $\mu_B$ is
close to the measured value\cite{garlea} of 1.3 $\mu_B$.
 Also, the calculated magnetic moment
at the Mn site is found to be 4.24 $\mu_B$ in good agreement
with the experimental estimate
\cite{garlea}.
 The orbital
moment at the V site seems to develop an appreciable value only beyond a critical U value,
U$_c$ (3.0 eV $<$ U$_c$ $\le$ 4.5 eV)\cite{orb}, which
may be interpreted as {\it Coulomb enhanced spin-orbit	effect}\cite{oka}.

We note that the perfect antiferro-orbital ordering as proposed by Refs. \onlinecite{suzuki} and
\onlinecite{garlea} would imply a quenching of orbital moment. The presence of a finite
orbital moment can be associated with the breakdown of perfect
 antiferro-orbital
ordering and may explain the domain alignment by magnetic field as observed by
Ref. \onlinecite{suzuki}.

We have also computed the magnetic exchange couplings from first principles by considering
LSDA+U total energy calculations with the PAW basis  for  different
spin alignments of V atoms within the V tetrahedra. Mapping the total energies to a Heisenberg
like model, we obtain  exchange interactions along the orbital chains (J) of 11 meV and
between the chains (J$^{'}$) of 2 meV. This implies $\alpha$ = J$^{'}$/J $\approx$ 0.2 compared to 0.3 found by
Chung {\it et al.}\cite{chung}. Perfect antiferro-orbital ordering with xz and yz alternately
occupied along the {\it c}-axis would however yield much smaller ratios of J$^{'}$/J, since the overlap
between orthogonal yz and xz orbitals at neighboring sites would have been nearly zero. The moderately
strong value of J$^{'}$, as obtained in the DFT calculation, originates from large mixing
of different t$_{2g}$ orbitals influencing the overlap of the renormalized orbitals at neighboring
sites.

Our calculations described so far assume the collinear arrangement of V spins, while experiment reports
a transition from collinear to noncollinear spin arrangements coincident with the structural phase transition.
In order to check whether our proposed orbital order sustains a
noncollinear arrangement of V spins, we
performed PAW  calculations where we relaxed the V spin orientation
 keeping the Mn spins aligned
parallel to the c axis\cite{garlea}. The relaxed spin
structure shows the V spins to be canted with respect to the c axis by about 63$^{\circ}$, which is
in very good agreement with the experimentally estimated canting of 65$^{\circ}$\cite{garlea}. The noncollinear
spin arrangement was found to be slightly favoured over the collinear ferrimagnetic spin arrangement
by an energy gain of 3 meV. Though this energy difference is almost within the accuracy limit
of DFT, the good agreement between theory and experimental estimates
is encouraging.

To conclude, we have carried out DFT-based first-principles calculations to investigate the nature of the orbital
ordering in \mn which is closely
associated with the transition from a high temperature cubic structure to a
low temperature tetragonal structure. Our geometry-optimized structures for \mn show a strong influence
of correlation effects in the choice of	 the {\it correct} low temperature structure.
The obtained ground state structure, $I4_1/a$ looses the mirror
and glide symmetry compared to the alternative proposed candidate $I4_1/amd$.
 The O in $I4_1/a$ are in 16f positions with nonzero x-coordinate, which makes the V-O
bondlengths even in the ab-plane to be unequal.	 This lowering of symmetry necessarily breaks the
degeneracy of the t$_{2g}$ states completely and also introduces mixing between different t$_{2g}$
states. The resulting eigenstates therefore turn out to be of {\it mixed-character} and {\it nondegenerate},
which get filled up by two V electrons. The occupied orbitals follow the site symmetry of vanadium which is
4-fold rotation times inversion to give rise to orbital chains with orbitals rotated with respect to each other
both within and between the chains.   Our DFT computed V-V magnetic coupling
 is found to be in agreement with the
experimental findings\cite{chung}. These results provide an explanation
of the controversy between  {\it antiferro-orbital ordering} versus the
strong exchange between orbital chains (J$^{'}$). We further showed
that our proposed orbital ordering  is capable of
predicting correctly the noncollinear spin structure as observed experimentally\cite{garlea}.
Further experiments like X-ray resonant spectroscopy would be helpful
to probe directly our proposed orbital order.

{\it Acknowledgements-}
We acknowledge useful discussions with
  J. Glinnemann and D. Khomskii. TSD thanks
Swarnajayanti Grant and MPI, Stuttgart through	partnergroup program.
RV thanks the DFG  for financial support through the SFB/TRR49
program. SS thanks CSIR for financial support.


\begin{thebibliography}{99}
\bibitem{note} Since Mn and V occupy the high symmetry 4a and 8d positions,
optimization essentially involves optimization of the internal degrees of freedom associated with O.
\bibitem{uvalue} Mn and V are
neighbors in the periodic table and it is not to be expected that their U values
will be very different.
\bibitem{double}   In all  calculations we considered the
 self interaction double counting correction (SIC)\protect\cite{Anisimov_93}.  We
also performed calculations with the around mean field correction (AMF) in order to check the influence of double counting
correction. Our conclusions
remained unchanged.   $I4_1/a$ is favored with respect to
 $I4_1/amd$ by 0.07 eV in the SIC calculation and 0.04 eV in the AMF calculation in APW+lo basis.
\bibitem{note1} The crystallographic $ab$-plane is rotated by 45$^{\circ}$ from the basal
plane defined by the V-O bonds, thereby forming t$_{2g}$ states
 of x$^{2}$-y$^{2}$, xz and yz symmetry
instead of more commonly used convention of xy, xz and yz.
\bibitem{orb} The V orbital moment computed for U values 3.0 eV, 4.5 eV and 6.0 eV were found to be 0.04 $\mu_B$,
0.34 $\mu_B$ and 0.38 $\mu_B$ respectively.
\end{thebibliography}
\end{document}